# DASP: Defect and Dopant ab-initio Simulation Package

Menglin Huang[1], Zhengneng Zheng[1], Zhenxing Dai[1], Xinjing Guo[1], Shanshan Wang[1], Lilai Jiang[1], Jinchen Wei[1], and Shiyou Chen[1,*]

[1]Key Laboratory of Computational Physical Sciences (MOE), and State Key Laboratory of ASIC and System, School of Microelectronics, Fudan University, Shanghai 200433, China

*chensy@fudan.edu.cn

**Abstract**

In order to perform automated calculations of defect and dopant properties in semiconductors and insulators, we developed a software package, Defect and Dopant ab-initio Simulation Package (DASP), which is composed of four modules for calculating: (i) elemental chemical potentials, (ii) defect (dopant) formation energies and transition energy levels, (iii) defect and carrier densities and (iv) carrier dynamics properties of high-density defects. DASP uses the materials genome database for quick determination of competing secondary phases and calculation of the energy above convex hull when calculating the elemental chemical potential that stabilizes compound semiconductors, so it can perform high-throughput prediction of thermodynamic stability of multinary compounds. DASP calls the ab-initio softwares to perform the total energy, structural relaxation and electronic structure calculations of the defect supercells with different structure configurations and charge states, based on which the defect formation energies and transition energy levels are calculated and the corrections for electrostatic potential alignment and image charge interaction can be included. Then DASP can calculate the equilibrium densities of defects and electron and hole carriers as well as the Fermi level in semiconductors under different chemical potential conditions and different growth/working temperature. For high-density defects, DASP can calculate the carrier dynamics properties such as the photoluminescence (PL) spectrum, defect-related radiative and non-radiative carrier capture cross sections, and recombination lifetime of non-equilibrium carriers. DASP is expected to act as an automatic and reliable toolbox for calculating the defect and dopant properties, which can be compared to or used to interpret the experimental characterization results of defects using electrical and optical techniques such as PL and deep level transient spectroscopy (DLTS). Here we will introduce its unique functions including the maximumly-cubic supercell generation, distorted defect structure searching and wavefunction initialization which can reduce computational cost and improve accuracy, and also show three examples about its applications in undoped GaN, C-doped GaN and quasi-one-dimensional SbSeI.

# 1. Introduction

Intrinsic point defects are indispensable in crystals at a finite temperature and can have important influences on the fundamental properties of crystalline materials.[1, 2] Extrinsic elements are usually introduced into the crystalline lattice as dopants intentionally or as impurities unintentionally, which provides an extrinsic freedom for manipulating the properties of crystalline materials. The influences of defects and dopants (impurities) in semiconductors are more significant than those in metals or structural materials, because the applications of semiconductors in functional electronic and optoelectronic devices care more about the electrical and optical properties, which can be significantly changed even when a small amount of defects and dopants exist in the lattice, *e.g.*, defects or dopants with a density of $10^{15}$-$10^{18}$ cm$^{-3}$ (1 defect or dopant among $10^4$-$10^7$ atoms) can change the electrical and optical properties of semiconductors dramatically. For the simple elemental semiconductors such as Si and Ge, the number of intrinsic point defects is usually small,[3] so extrinsic doping is required for producing electron or hole carriers and thus achieving n-type or p-type electrical conductivity. For binary, ternary, quaternary or even multinary compound semiconductors, the number of possible point defects can be large, so their defect/dopant physics can be more complicated and thus more flexible, *e.g.*, their electrical conductivity may be tuned through controlling the formation of different point defects.[4] Considering the important influences of defects (dopants) on semiconductor properties and the complexity of the defect/dopant physics in new, multinary or low-symmetry semiconductors, the studies on defects and dopants are fundamental to the development of semiconductor physics and applications.

Many experimental techniques have been developed for the characterization of defects and dopants in semiconductors, such as the photoluminescence (PL) spectrum, positron annihilation spectroscopy (PAS), electron paramagnetic resonance (EPR) and deep-level transient spectroscopy (DLTS).[5] These techniques usually just give the measured spectra from which the defect and dopant properties such as the density, energy levels and carrier capture cross sections can be obtained through fitting the spectra according to a certain physical model. However, what kind of defects and dopants determine the experimental spectra (including the fitted properties of defects and dopants) is usually speculated according to the analysis on the properties of the possible defects and dopants, which is semi-empirical and imposes a limit to the accurate and quantitative defect and dopant engineering. *For instance*, if the origin defects of the deep-level recombination centers in optoelectronic semiconductors are not clear, it is impossible to suppress their formation and increase the minority carrier lifetime by accurately controlling the growth conditions.

Since 1990s, significant progresses have been made in the ab-initio (first-principles) calculation of defect and dopant properties based on the density functional theory (DFT) and the supercell model.[6-8] The calculated formation energies, charge-state transition energy levels and carrier capture cross sections of defects and dopants have been widely used for revealing the microscopic origin of the experimental PL, PAS, EPR and DLTS spectra, and guiding the fabrication of high-quality materials and high-performance devices.[9-11] However, there are still unsolved issues that may cause considerable errors in the calculated defect/dopant properties or even misunderstanding of the defect/dopant physics. (i) The approximations to exchange-correlation functionals cause the underestimation of the band gap and the incorrect location and occupation of the defect/dopant levels in the band gap,[12] which can result in errors in the calculated formation energy and transition level. Such errors are serious for the local density approximation (LDA)[13] and generalized gradient approximation (GGA)[14], so several correction schemes were proposed.[7, 15-18] Recently, the hybrid functional calculations were shown to predict more reasonable band gaps for many semiconductors, so the errors in the calculated defect properties are also corrected largely (not perfectly, because the hybrid functional may give incorrect localization of defect states and also depends on the empirical exact exchange ratio parameter).[19-21] (ii) The calculation of defect/dopant formation energies and densities requires the determination of the allowed chemical potential ranges for all the component and dopant elements. For ternary, quaternary and multinary compound semiconductors, there probably exist a large number of secondary phases competing against the host material, which may all limit the stable range of elemental chemical potentials. Therefore, if some important secondary phases are not considered in the determination of elemental chemical potentials, the calculations of defect/dopant formation energies and densities and even the stability of the compound can be wrong. (iii) For low-symmetry and multinary compound semiconductors, the possible defect/dopant types can be many and various. For each type of defect/dopant, there may be multiple non-equivalent sites and structural configurations, which can change for different charge states. If some important defect-types/sites/configurations/charge-states are not considered, the calculated results are incomplete and inaccurate. (iv) The finite supercell size corrections, such as the corrections for the electrostatic potential alignment and image charge interaction,[12, 22, 23] are still controversial. Especially, different correction schemes can give rise to different results. (v) In most of the studies, only defect/dopant formation energies and transition levels were calculated, but the exact values of equilibrium defect densities, carrier densities and Fermi level are not calculated. Without the direct calculation of the defect densities, one may have misunderstandings on the influences of a certain defect, because all the defects (dopants) are correlated by the carrier densities and Fermi level, and the defect correlation may give rise to unexpected and anti-chemical-intuition changes in the defect densities.[24]

To solve those problems in a standard and comprehensive way, we developed a software package, Defect and Dopant ab-initio Simulation Package (DASP), which can perform automated calculations of defect and dopant properties based on the supercell model and ab-initio DFT calculations. Using the software, the defect formation energies, charge-state transition energy levels, carrier capture cross sections and even the PL spectrum can be calculated based on the atomic structure, total energy, electronic structure, phonon spectrum and electron-phonon coupling matrix calculated by the ab-initio DFT softwares using different approximations to exchange-correlation functionals. All the possible competing secondary compounds in the materials genome database are considered for the accurate calculation of the thermodynamic stability and elemental chemical potential ranges of compound semiconductors. Various defect types, atomic sites, structure configurations and charge states can be considered, and the corrections for the electrostatic potential alignment and image charge interaction can be included automatically. With all the defects and dopants considered, the equilibrium defect densities, carrier densities and Fermi level can be calculated for the samples grown under different chemical potential conditions and different temperature. For high-density defects and dopants, their carrier dynamics properties can also be calculated, despite the heavier computational cost. Therefore, DASP is developed to be a toolbox for the automated theoretical prediction of defect and dopant properties in semiconductors.

## 2. Software Framework and Modules

### 2.1 Software Framework

DASP is composed of four modules: Thermodynamic Stability Calculation (TSC), Defect Energy Calculation (DEC), Defect Density Calculation (DDC), and Carrier Dynamics Calculation (CDC), as plotted in Fig. 1.

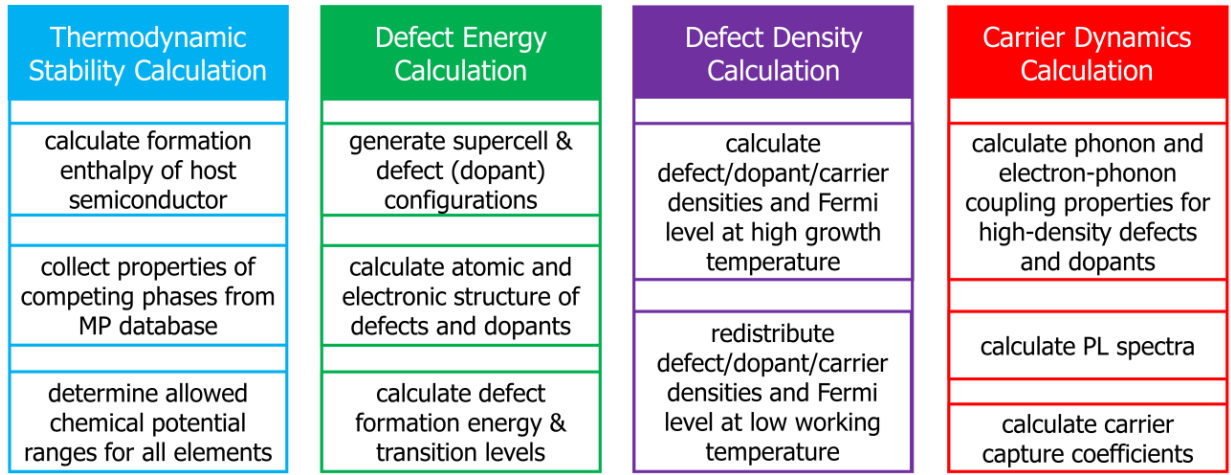

Figure 1. The framework of the DASP software, which is composed of four modules, TSC, DEC, DDC and CDC. The major functions of the four modules are shown in the boxes.

The detailed workflow of DASP is plotted schematically in Fig. 2, in which the green part shows the DEC module, the blue part shows the TSC module, the purple part shows the DDC module and the red part shows the CDC module. The only necessary input is the crystal structure file of the semiconductor. The intermediate and final output files include:

(i) TSC: the chemical potential range of component elements that stabilizes the compound semiconductor (which is also a descriptor of the thermodynamic stability of the compound) and the allowed highest chemical potential of the dopant elements;

(ii) DEC: the formation energies of defects and dopants in different charge states, as functions of the elemental chemical potentials and Fermi level (electronic chemical potentials), from which the charge-state transition energy levels can be derived;

(iii) the equilibrium-state Fermi level, densities of electron and hole carriers, and densities of the defects and dopants in different charge states, as functions of the elemental chemical potentials at growth temperature and working (measuring) temperature;

(iv) the carrier capture cross sections, the radiative and non-radiative carrier recombination rates and lifetime, and the PL spectra.

Among the four modules, the most fundamental one is DEC, whose core function is for calculating the defect/dopant formation energy using the supercell model and ab-initio DFT calculations. In the supercell model, the formation energy of a point defect $\alpha$ in the charge state $q$ can be calculated as,[25, 26]

$$\Delta E_f(\alpha, q) = E(\alpha, q) - E(bulk) - \sum_i n_i(\mu_i + E_i) + q(E_F + \varepsilon_{VBM}) + E_{corr} \quad (1),$$

where $E(\alpha, q)$ and $E(bulk)$ are the total energies of the supercells with and without a defect (dopant), $\mu_i$ is the elemental chemical potential referenced to the total energy $E_i$ of the pure solid/gas elementary phase, $E_F$ is the Fermi level referenced to the eigenvalue of the valence band maximum (VBM) level of the bulk supercell. $E_{corr}$ is the correction that accounts for the spurious interaction caused by finite supercell size and periodic boundary conditions.

In the following, we will briefly introduce the four modules:

**TSC Module:** As shown in Eq. (1), the defect (dopant) formation energy and thus the defect density are functions of the elemental chemical potentials, so the calculation of defect properties requires the elemental chemical potentials as inputs. The TSC (Thermodynamic Stability Calculation) module is for calculating the chemical potential ranges through considering the influences of all the competing secondary compounds that can limit the thermodynamic phase stability of the host compound semiconductor. In order to make the consideration of secondary compounds as complete as possible, the material genome database (Materials Project[27]) is used to search for the competing secondary compounds. Through combining the formation energy information of all the secondary compounds with the calculated formation energy of the host compound (equivalent inputs are used for DFT calculations), the critical secondary compounds that limit the stable chemical potential region can be determined quickly.

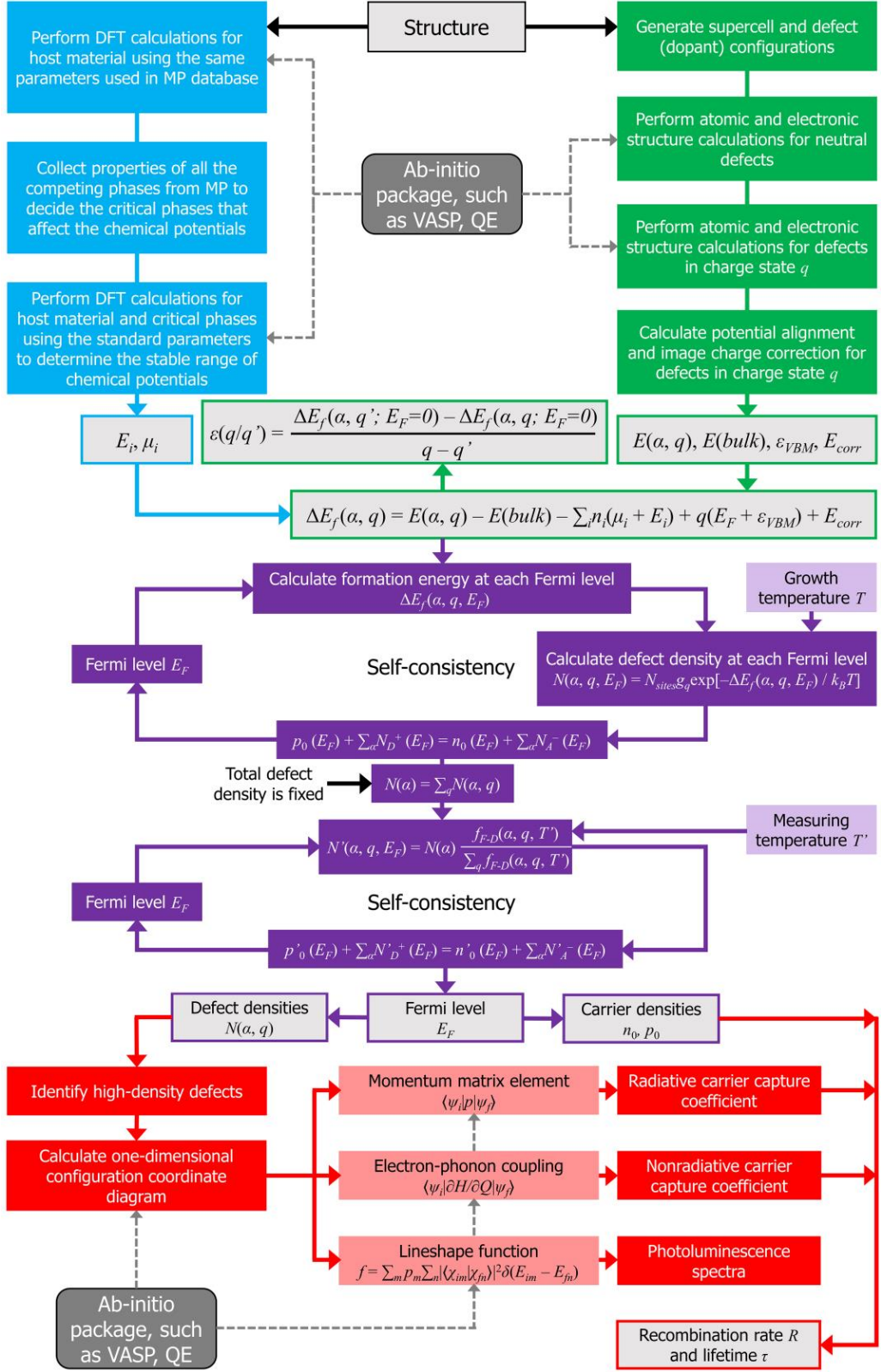

Figure 2. The flowchart of DASP. Different colors represent the four modules. The dashed lines show the calculations that need to call external ab-initio DFT softwares.

**DEC Module:** The DEC (Defect Energy Calculation) module can generate the maximumly-cubic supercell structures of the host semiconductor, based on which the defect and dopant supercell structures can be generated with all the possible defect/dopant types, atomic sites and structural configurations considered. Then ab-initio softwares, such as Vienna Ab-Initio Simulation Package (VASP)[28] and Quantum ESPRESSO (QE)[29], will be called to perform atomic structural relaxation, total energy and electronic structure calculations for the defect/dopant supercells. First the neutral charge state is calculated and then different charge states will be generated according to the defect/dopant levels in the band gap and calculated. Then the formation energies of defects and dopants in different charge states can be calculated using Eq. (1). The chemical potentials outputted by TSC are used as $\mu_i$. The correction term $E_{corr}$ that accounts for the spurious interaction caused by finite supercell size and periodic boundary conditions can also be calculated automatically based on the output of ab-initio calculations. The charge-state transition energy levels are then calculated to be the Fermi level $E_F$ at which the formation energies of a defect in two charge states $q$ and $q'$ are equal, $\Delta E_f(\alpha, q) = \Delta E_f(\alpha, q')$.

**DDC Module:** Based on the results of TSC and DEC, the DDC (Defect Density Calculation) module solves the charge-neutrality equation self-consistently to determine the equilibrium defect/dopant densities, Fermi level, and carrier densities, as shown by the purple part of Fig. 2. For a given chemical potential condition, the formation energies of defects and dopants in different charge states are treated as input for a two-step self-consistent calculation, which solves the charge-neutrality condition equations at high growth temperature and low working (measuring) temperature to determine the Fermi level of the semiconductor samples. Then the equilibrium densities of defects and dopants and the densities of electron and hole carriers can be calculated under the given chemical potential condition. Through changing the chemical potentials, the defect/dopant and carrier densities can be calculated as functions of the chemical potential conditions, which can be used for the defect/dopant engineering based on the growth condition control.

**CDC Module:** For the high-density defects or dopants identified by DDC module, the CDC (Carrier Dynamics Calculation) module can calculate their excited-state carrier dynamics properties. The transition energy levels from the DEC module and the Fermi level from the DDC module will be used to screen for the transitions between the defect levels and the VBM or conduction band minimum (CBM) level which may produce the PL peaks or cause the fast non-radiative recombination of carriers. Then the phonon spectrum of the defect/dopant supercell, electron-phonon coupling matrix and transition dipole moments between the defect/dopant states and VBM or CBM states will be calculated, based on which the radiative and non-radiative transition rates (carrier capture cross sections) and the lineshape of photoluminescence spectra induced by defects and dopants can be calculated. Such calculations may be performed based on the single-phonon-mode one-dimensional configuration coordinate diagram[30] or the all-phonon-modes three-dimensional schemes[31]. Combining the calculated transition rates and the equilibrium defect and carrier densities, the CDC module may also calculate the Shockley-Read-Hall recombination rates and the non-equilibrium carrier lifetime.

## 2.2 Thermodynamic Stability Calculation Module

The formation of defects or the doping of extrinsic elements in semiconductor lattice involves the exchange of atoms between the semiconductor lattice and the external environment, so the formation energies of defects and dopants in Eq. (1) depend on the elemental chemical potentials, which describe the abundance of the elements (partial pressure for the gas elements) in the environment. Although the abundance of a certain element can be controlled by changing the environment, the changes of the elemental chemical potentials are actually not unlimited, because they should satisfy a series of thermodynamic conditions to make the pure-phase crystalline semiconductor stable. Here we take the quaternary compound semiconductor $Cu_2ZnSnS_4$ as an example to show how to determine its stable elemental chemical potential range.

Firstly, at the equilibrium state of the compound $Cu_2ZnSnS_4$, the weighted sum of the chemical potentials of its component elements should be equal to the formation energy of the compound,

$$2\mu_{Cu} + \mu_{Zn} + \mu_{Sn} + 4\mu_S = \Delta H_f(Cu_2ZnSnS_4) \quad (2),$$

where $\Delta H_f(Cu_2ZnSnS_4)$ is the compound formation energy of $Cu_2ZnSnS_4$ relative to the Cu, Zn, Sn and S elemental phases. That means the chemical equilibrium state is arrived for the reaction $2Cu + Zn + Sn + 4S \rightarrow Cu_2ZnSnS_4$, so the product compound $Cu_2ZnSnS_4$ is stable.

Secondly, to make the synthesized sample be pure-phase $Cu_2ZnSnS_4$, the formation or coexistence of the competing secondary phases, such as the binary and ternary compounds CuS,

$Cu_2S$, $ZnS$, $SnS$, $SnS_2$ and $Cu_2SnS_3$, and the elemental phases Cu, Zn, Sn, S, should be avoided. Therefore, the weighted sum of the chemical potentials of their component elements should be lower than their corresponding formation energies (the formation energies of elemental phases are 0), as described by the following inequations,

$$\mu_{Cu} + \mu_S < \Delta H_f(CuS),$$

$$2\mu_{Cu} + \mu_S < \Delta H_f(Cu_2S),$$

$$\mu_{Zn} + \mu_S < \Delta H_f(ZnS),$$

$$\mu_{Sn} + \mu_S < \Delta H_f(SnS),$$

$$\mu_{Sn} + 2\mu_S < \Delta H_f(SnS_2),$$

$$2\mu_{Cu} + \mu_{Sn} + 3\mu_S < \Delta H_f(Cu_2SnS_3),$$

$$\mu_{Cu} < 0,$$

$$\mu_{Zn} < 0,$$

$$\mu_{Sn} < 0,$$

$$\mu_S < 0 \quad (3).$$

The allowed chemical potential range of Cu, Zn, Sn and S that stabilizes the pure-phase $Cu_2ZnSnS_4$ is limited by these equations and inequations. If the formation energies of the host and all the competing compounds are known, the ranges can be calculated and plotted in the chemical potential space, as shown in Ref. [4]. One example about SbSeI is also shown in Section 3.2.

The four elements Cu, Zn, Sn and S can form many binary, ternary and quaternary compounds, which can all act as the competing secondary phases of $Cu_2ZnSnS_4$. In principle, for the accurate calculation of the chemical potential range, one should take account of all the possible competing compounds, whose number can be large, especially for quaternary or multinary compounds. In DASP, the TSC (Thermodynamic Stability Calculation) module provides an automated solution for the quick screening of the critical competing phases and thus the accurate calculation of the chemical potential range. With the chemical formula of the compound, TSC will visit the materials genome database, such as Materials Project (MP) database[27], to search for all the compounds that are composed of the component elements, and then download the total energy, formation energy and DFT calculation input files of these compounds. Meanwhile, TSC will also perform ab-initio DFT calculation for the formation energy of the host compound with the input consistent with the MP database. With the formation energies of the host compound and all the competing compounds, TSC can solve the thermodynamic constraint equations and inequations to determine the stable chemical potential region and the critical competing compounds that limit the stable region directly. For the host and the critical competing compounds, TSC can further calculate their formation energies with higher accuracy and different functionals, to ensure that the calculated range of the chemical potentials are accurate. Usually these critical competing compounds are also those determining the energy above the convex hull, which shows the thermodynamic stability of the compound with respect to the phase separation into other competing compounds, so TSC can also be used for high-throughput and accurate calculation of the energy above convex hull and thermodynamic stability of compounds.

### 2.3 Defect Energy Calculation Module

### 2.3.1 Maximumly-Cubic Supercell Generation

In the real semiconductor lattices, the densities of defects and dopants are usually much lower than the densities of atoms, *e.g.*, there is only one defect or dopant among $10^4$-$10^7$ atoms for defects or dopants with a density of $10^{15}$-$10^{18}$ cm$^{-3}$. Therefore, the distance between two defects or dopants is usually very large. However, in the periodic supercell model, the distances between the defect and the periodic image defects in the neighboring supercell are actually not very large since the supercell has only several hundred atoms. Therefore, the supercell model causes unphysical interaction between defects or dopants, which may induce errors in the calculated defect properties.

To reduce the errors caused by the finite supercell size and achieve better supercell size convergence, DASP always tends to generate the supercells that are nearly cubic, so that the defect-defect (dopant-dopant) distance is maximized and the unphysical interaction is minimized. As illustrated in Fig. 3, for the two supercells with the same volume (same number of atoms), the smallest defect-defect distance in the nearly-cubic supercell should be larger than that in the supercell derived from direct expansion of primitive cell. In the nearly-cubic supercell, the smallest defect-defect distance should be along the (100), (010) or (001) direction, and the smallest distances along the three directions are similar. However, in the supercell derived from direct expansion of primitive cell, the smallest defect-defect distance may be along the (110) or (111) directions, and the distance can be smaller than the distance along the (100), (010) or (001) direction if the supercell basis vectors are not orthogonal and one of their angles is larger than 120°. Besides the advantage in maximizing the defect-defect distance, the nearly-cubic supercell also leads to the orthogonality of the reciprocal lattices, which may accelerate the ab-initio calculation and achieve better convergence.

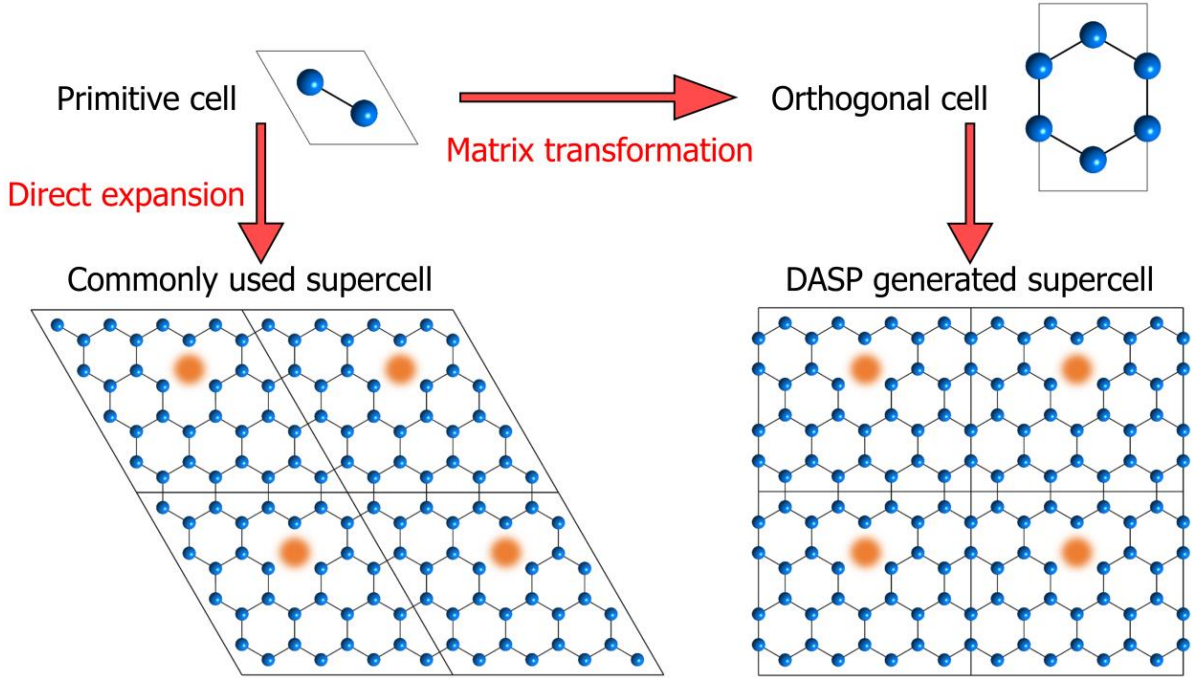

Figure 3. The supercell generated by simple expansion of primitive cell and the maximumly-cubic supercell generated by DASP.

The generation of the nearly-cubic supercell in DASP is through the maximumly-cubic supercell generation function, which firstly transforms the primitive cell into the nearly-orthogonal supercell and then expands it to get the nearly-cubic supercell which maximizes the cubic degree of the supercell with a given number of atoms.

To search for the nearly-orthogonal supercell, we developed a traversal searching method through trying different conversion matrices that can transform the lattice vectors of the primitive cell to the nearly-orthogonal lattice vectors. The conversion matrix should be non-singular because the lattice vectors are not coplanar. In addition, only swapping the lattice vectors will not change the shape of the cell, so there is no need to consider the exchange operation during the transformation. Therefore, the matrix satisfies the LU decomposition condition, and can be written as,

$$C = \begin{pmatrix} c_{11} & c_{12} & c_{13} \\ c_{21} & c_{22} & c_{23} \\ c_{31} & c_{32} & c_{33} \end{pmatrix} = LU = \begin{pmatrix} 1 & 0 & 0 \\ l_{21} & 1 & 0 \\ l_{31} & l_{32} & 1 \end{pmatrix} \begin{pmatrix} u_{11} & u_{12} & u_{13} \\ 0 & u_{22} & u_{23} \\ 0 & 0 & u_{33} \end{pmatrix} \quad (4),$$

where $u_{11}, u_{22}, u_{33}$ in the upper triangular matrix $U$ describe the expansion operations, which are the repeating numbers along the three lattice vectors of the primitive cell. $l_{21}, l_{31}, l_{32}$ in the lower triangular matrix $L$ and $u_{12}, u_{13}, u_{23}$ in the upper triangular matrix $U$ describe the linear combination operations. By traversing different conversion matrices, we can get different nearly-orthogonal cells. Taking into account both the computational cost and accuracy, we only traverse the conversion matrices that meet the following conditions,

$$c_{ij} = 0, \pm 1, \pm 2, \pm 3, \pm 4 \ (i, j = 1, 2, 3), \quad u_{kk} = 1, 2, 3, 4 \ (k = 1, 2, 3) \quad (5),$$

where $c_{ij}$ is the element in the conversion matrix $C$, $u_{kk}$ is the diagonal element in the upper triangular matrix $U$.

With the nearly-orthogonal supercell, we can just expand it directly to get nearly-cubic supercell. The cubic degree is described by the ratio $\frac{V_{supercell}}{V_{cube}}$, where $V_{supercell}$ is the volume of the supercell, $V_{cube}$ is the volume of the cube that can envelop the whole supercell. If the degree is 1, the shape of the supercell is ideally cubic. Considering the computational cost, the number ($N_{atom}$) of atoms in the supercell is limited, e.g., 200-1000 atoms. Although very large $N_{atom}$ can give very cubic supercell and thus tend to increase the cubic degree to 1, the cubic degree does not necessarily increase monotonically as $N_{atom}$ increases from 100 to 1000. Therefore, when selecting a supercell for defect calculation, DASP considers both the cubic degree and $N_{atom}$ in order to maximize the defect-defect distance under the constraint of the limited $N_{atom}$. An empirical quantity $\beta$ is defined to balance the two factors and used as the criterion for choosing supercells,

$$\beta = \frac{V_{supercell}}{V_{cube}} + 0.001 \, N_{atom} \quad (6).$$

The maximumly-cubic supercell is generated by DASP through maximizing the quantity $\beta$ in a given range of $N_{atom}$.

**2.3.2 Distorted Defect Structure Searching**

After the supercell is generated, DASP can generate structures of defects and dopants in the primitive-cell region of the supercell. The common defect types will be considered automatically. For example, the intrinsic defects of the quaternary compound $Cu_2ZnSnS_4$ that DASP considered include the Cu, Zn, Sn and S vacancies ($V_{Cu}$, $V_{Zn}$, $V_{Sn}$ and $V_S$), interstitials ($Cu_i$, $Zn_i$, $Sn_i$, $S_i$) and antisites ($Cu_{Zn}$, $Cu_{Sn}$, $Cu_S$, $Zn_{Cu}$, $Zn_{Sn}$, $Zn_S$, $Sn_{Cu}$, $Sn_{Zn}$, $Sn_S$, $S_{Cu}$, $S_{Zn}$, $S_{Sn}$). Meanwhile, for the low-energy donor and acceptor defects, they can form donor-acceptor compensated defect complexes, which can also be considered by DASP.

For low-symmetry semiconductors, the structure may have several non-equivalent atomic sites for one element, and the same-type defects on non-equivalent sites may have different properties. When generating vacancy, interstitial and antisite defects, DASP will consider all non-equivalent sites. For interstitial defects, DASP will search for the largest void region in the structure and meanwhile consider the Coulomb repulsion to determine the possible interstitial sites of cations and anions. For low-energy interstitial defects, DASP will also generate 10-20 different configurations randomly in the primitive-cell region of the supercell (different interstitial sites are ensured to be not close to each other).

For low-energy vacancy and antisite defects, there may be other distorted or metastable structure configurations, such as the DX centers[32] which originate from large structural distortion of the antisite donor defects but act as acceptors after distortion. These distorted or metastable defects have been shown to have important impact on the electrical properties of semiconductors.[33-35] They can be obtained by imposing a structural perturbation to the locally relaxed structure of the defect in the charge state $q$, because the perturbation may overcome the structural transition barrier to arrive at a new distorted structure.

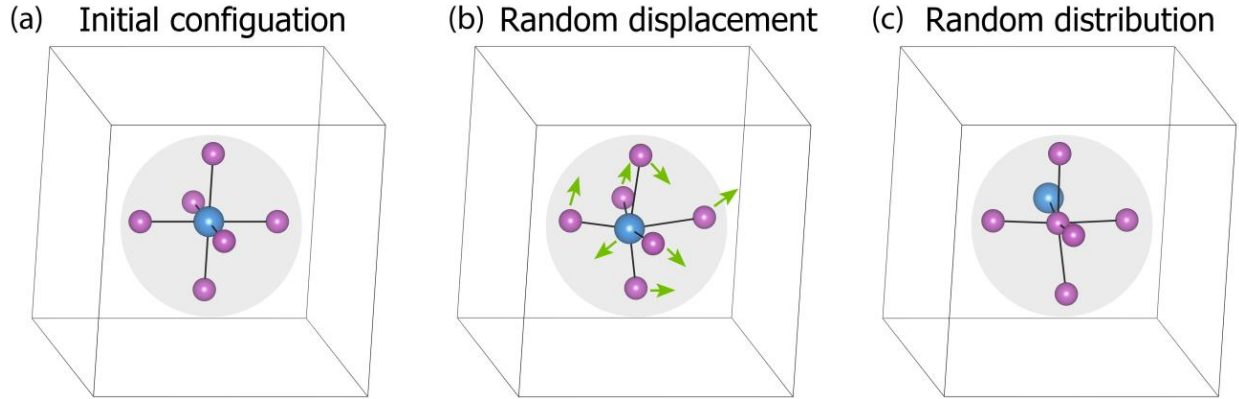

Figure 4. For a vacancy or antisite defect, the initial configuration (a) can be generated directly from the host lattice, and then structural perturbations are added, including (b) random displacements and (c) random distribution of the atoms within the sphere around the defect.

DASP adds two types of structural perturbations: (i) displacing the atoms within a sphere around the defect by less than 0.5 Å randomly, as shown in Fig. 4(b); (ii) randomly distributing all the atoms within the sphere, as shown in Fig. 4(c). The default value of the sphere radius is set to 3 Å, but will be automatically increased to ensure at least 4 atoms in the sphere. The total number of the structures generated with the perturbation will be in the range of 10-20, but some of them may become identical after structural relaxation. Such kind of structural perturbations are in fact for achieving a global structural relaxation for the defect.

### 2.3.3 Charge State Selection of Ionized Defects

Ionized defects in different charge states are fundamental to the basic understanding of defect physics, but there are still several open questions during the calculation of the properties of charged defects. The first one is how to determine the range of defect charge $q$ for an ionized defect. In the past, the octet rule was often selected as a criterion to judge the charge range of a defect, which simply adopts the nominal charge of the single element. For instance, the sulfur vacancy should take 0, 1+, 2+ charges according to the simple octet rule, however, in non-conventional semiconductors such as $MoS_2$, neither 1+ nor 2+ state of the sulfur vacancy is stable while the 1– charge state exists.[36]

To solve this problem, the estimation procedure of the charge range implemented in DASP is based on the calculated eigenvalues of the neutral defects at Γ point. Therefore, in the first step of DEC module, DASP will generate the structures of point defects in their neutral state, and then call ab-initio software to perform atomic relaxations and static self-consistent calculations. Once all the calculations are done, the eigenvalues of bulk and all the neutral defects will be extracted. The hybrid functional (such as HSE) is recommended for the static self-consistent calculations to obtain the more reasonable band gap and more reasonable location of the defect levels. As shown in Fig. 5, for defect A, if an occupied and an unoccupied level are found within the band gap between the VBM and CBM levels of the bulk, these two levels will be taken as defect levels and the defect A in 1+ and 1– charges will be calculated subsequently. Following this scheme, when three occupied levels and three unoccupied levels are found in the bulk band gap, the defect C in [3+, 2+, 1+, 1–, 2–, 3–] charge states will all be calculated.

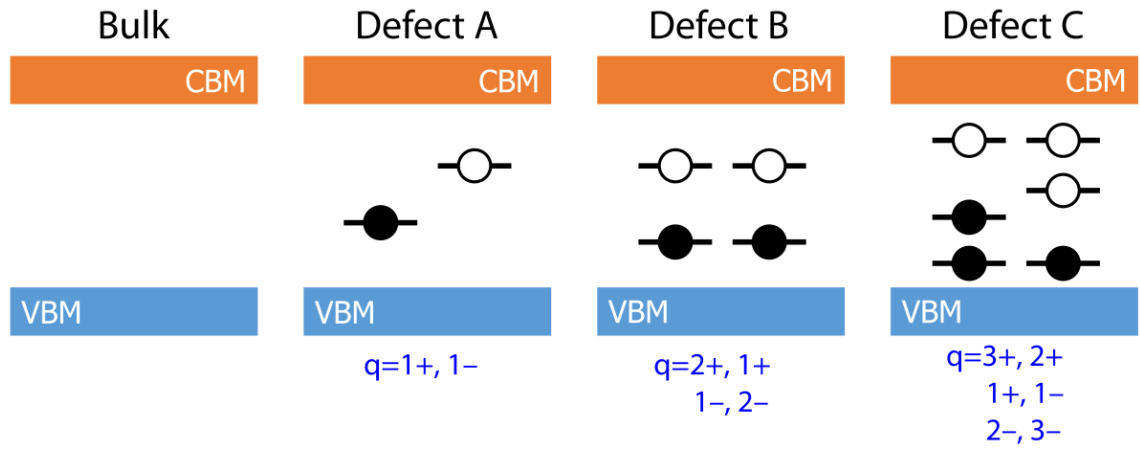

Figure 5. The determination of the charge states of the ionized defect according to the calculated eigenvalues of the defect levels within the bulk band gap, extracted from the ab-initio calculation for the neutral defect.

### 2.3.4 Electrostatic Potential Alignment

In Eq. (1), the eigenvalue of bulk VBM is used as the reference of the Fermi level, *i.e.*, $E_F=0$ means the Fermi level is located at VBM. However, the eigenvalues in the calculation of bulk and defect supercells can be compared only when the electrostatic potentials of the two periodic supercells are aligned.[12] This can be accomplished either by aligning the electrostatic potential far from the defect (*e.g.*, the potential written in the LOCPOT file of VASP), or by aligning the deep core level of the farthest atom from the defect. The alignment term can be written as,

$$\Delta V_{q/b} = V(\alpha, q)|_{far} - V(bulk)|_{far} \quad (7).$$

The correction of potential alignment to the formation energy of a defect in the charge state $q$ is $q\Delta V_{q/b}$, and is automatically included in the term $E_{corr}$.

### 2.3.5 Image Charge Correction

The spurious Coulomb interaction between charged defect and its periodic images, as well as charged defect and the neutralizing background charge should be corrected from the calculated formation energies of ionized charged defects, which are called image charge corrections. DASP can currently perform the corrections according to two schemes. The first one is the Lany-Zunger scheme,[12] which can be represented by,

$$\Delta E_{LZ} = [1 - c_{sh}(1 - 1/\varepsilon)]\frac{q^2\alpha}{2\varepsilon L} \quad (8),$$

where $\alpha$ is the Madelung constant, $\varepsilon$ is the static dielectric constant, and $L$ is the linear supercell dimension. Since DASP will always tend to generate a nearly-cubic supercell, the default value of $c_{sh}$ is set to –1/3.[12, 22] Based on these definitions, the total correction term $E_{corr}$ appears in Eq. (1) can be separated into,

$$E_{corr} = q\Delta V_{q/b} + \Delta E_{LZ} \quad (9).$$

DASP also provides an interface with the code *sxdefectalign* written by Freysoldt, which implements the Freysoldt-Neugebauer-Van de Walle (FNV) scheme[23] for charged defect correction. Different from the original paper that uses the planar average of the electrostatic potential to obtain the potential offset, we here use the atomic site potential instead, as Kumagai and Oba proposed.[37] In this scheme, the correction $E_{corr}$ can be written as,

$$E_{corr} = E_{lat} - q(\Delta V_{q/b} - V_{model}) \quad (10),$$

where $E_{lat}$ is the Madelung energy and $V_{model}$ is the potential obtained from the chosen model (such as Gaussian) charge density. Note that in the FNV scheme, the potential alignment defined in 2.3.4 is already included.

The two correction schemes mentioned above may be not valid for the charged defects in low-dimensional layered semiconductors, such as those in monolayer $MoS_2$ or $WSe_2$. A series of new correction schemes have been successfully developed in the past decade for the charged defect correction in layered semiconductors,[38-43] which will be implemented in the future version of DASP.

**2.3.6 Defect Wavefunction Initialization**

The formation of a defect or dopant causes the change of the electronic wavefunction and the redistribution of charge density in the semiconductor lattice, which produces new electronic states (defect states) in the band gap. Although the wavefunction and charge density near the defect site can be changed dramatically, the area far from the defect should be weakly influenced. Therefore, for different defects or dopants in different charge states, the wavefunction and charge density in the region far from the defect site should be similar to those of the defect-free supercell. Since the self-consistent calculations of the wavefunction and charge density will be repeated hundreds of times, it will save a large amount of computational cost if we take advantage of the slightly changed wavefunction and charge density in the region far from the defect site to reduce the number of self-consistent calculation steps.

In DASP, the initial charge density of the neutral defect/dopant supercell is generated based on the converged wavefunction and charge density of the bulk host supercell, *i.e.*, the charge density in the region far from the defect site is the same as that of the bulk host, while the charge density in the sphere around the defect is generated by a superposition of atomic charge densities and the charge density in the interface region is smoothed and rescaled according to the number of valence electrons of the defect supercell. Then the self-consistent calculations are performed to obtain the converged charge density and wavefunction of the neutral defect/dopant supercell. For the charged defect/dopant, its initial charge density is generated based on the converged wavefunction and charge density of the neutral defect through changing the occupation number of the defect states. With the initial charge density, the ab-initio self-consistent calculations can reach convergence with much less steps. The saving of computational cost can be more significant when large supercells are used.

## 2.4 Defect Density Calculation Module

Under thermodynamic equilibrium, the density of a defect $\alpha$ in the charge state $q$ is determined by its formation energy according to,[25, 44]

$$n(\alpha, q) = N_{sites} g_q e^{(-\Delta E_f/k_B T)} \quad (11),$$

where $N_{sites}$ is the density of the possible defect sites, $g_q$ is the charge-dependent degeneracy factor, $\Delta E_f$ is the defect formation energy at the given Fermi level and elemental chemical potentials as described by Eq. (1). All the ionized defects in the charge state $q \neq 0$ produce carriers. The positively charged donor defects with $q>0$ produce electron carriers, and their summed charge is $\sum_{\alpha;q>0}[q * n(\alpha, q)]$; while the negatively charged acceptor defects with $q<0$ produce hole carriers, and their summed charge is $\sum_{\alpha;q<0}[q * n(\alpha, q)]$. The final densities of electron and hole carriers are contributed by both the thermal excitation and the ionization of all these defects (dopants). The equilibrium-state Fermi level can be calculated through solving the charge neutrality equation,

$$n_0 + \sum_{\alpha;q<0}[q * n(\alpha, q)] = p_0 + \sum_{\alpha;q>0}[q * n(\alpha, q)] \quad (12),$$

where $\sum_{\alpha;q<0}[q * n(\alpha, q)]$ and $\sum_{\alpha;q>0}[q * n(\alpha, q)]$ are the summed charges of negatively charged defects and positively charged defects, weighted by the charge $q$. $n_0$ and $p_0$ are free carrier densities, which can be defined as,

$$n_0 = \int_{\varepsilon_{CBM}}^{+\infty} g_C(E) f(E) dE \quad (13),$$

$$p_0 = \int_{-\infty}^{\varepsilon_{VBM}} g_V(E)(1 - f(E)) dE \quad (14),$$

where $g_C(E)$ and $g_V(E)$ are the density of states (DOS) for the conduction and valence bands, respectively, and $f(E)$ is the Fermi-Dirac occupation function. $g_C(E)$ and $g_V(E)$ can be calculated from both the parabolic band approximation and the exact integration of the ab-initio calculated band structure. The default setting in DASP is from the parabolic band approximation which requires the calculation of electron and hole effective masses, but it can be changed to the exact DOS from ab-initio calculations.

The semiconductors are usually grown or synthesized at a high temperature and then go through a rapid annealing process to a lower working (measuring) temperature. Therefore, the defects are usually formed at the high temperature and then the densities of different charge states will redistribute during the rapid annealing. The DDC module is developed in accordance with such fabrication process,[44-46] as shown schematically in Fig. 2. Eq. (12) is firstly solved at a high growth temperature, and then the Fermi level and the densities of each defect in different charge state $q$ can be obtained at the high temperature. Afterwards, Eq. (12) is solved again at the lower measuring temperature, but now the densities of defects in different charge states do not follow Eq. (11). In the second step, the density summation for each defect over all charge states is fixed at the value calculated in the first step, and the density of each charge state will undergo a redistribution according to the Fermi-Dirac occupation. Then a new Fermi level can be obtained at the measuring temperature, and the redistributed defect densities and carrier densities can be calculated.

The calculation of defect and carrier densities using DDC module in DASP is quite easy. After finishing the calculation of TSC and DEC, DDC can be calculated based on the output files of TSC and DEC. The defect (dopant) densities, carrier densities and Fermi level can be plotted or saved as functions of the elemental chemical potentials and growth/measuring temperatures.

**2.5 Carrier Dynamics Calculation Module**

After calculating the defect densities using DDC module, one may identify a portion of high density defects (dopants), which may be critical to the optical and electrical properties of the host semiconductor. For those important defects, we implement in the CDC module three functions for studying the excited-state carrier dynamics properties based on the phonon spectrum and electron-phonon coupling calculation: (i) photoluminescence (PL) lineshape of defects; (ii) radiative carrier capture coefficient of defects; (iii) phonon-assisted nonradiative carrier capture coefficient (cross section) of defects. In the following, we will mainly introduce how to calculate the PL lineshape, while the details of the calculation on carrier capture will not be discussed here. More details can be found in Ref. [30, 31].

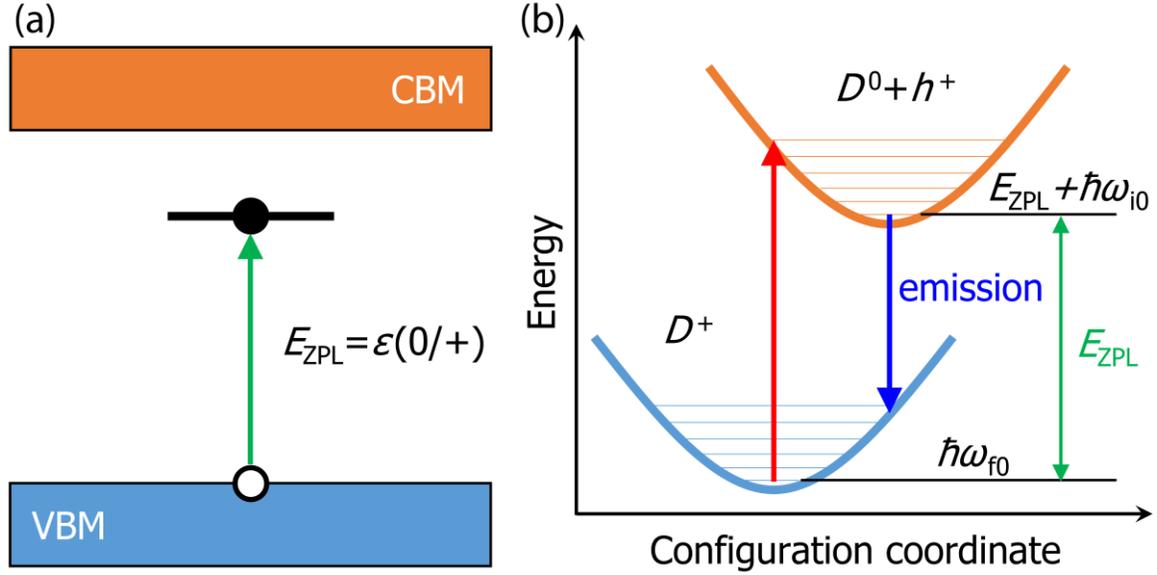

Figure 6. (a) Hole capture process by the donor defect D that changes the charge state from neutral to +1. (b) Configuration coordinate diagram of hole capture process. The potential curves are aligned to ensure the zero-phonon line energy equals to the (0/+) transition energy.

The photoluminescence can be caused by the transition of carriers between the defect state and the VBM or CBM state, as shown in Fig. 6(a). Its intensity at finite temperature is mainly determined by the transition dipole moment between two states and the lineshape function, which is written by,[47, 48]

$$I(\hbar\omega) = \frac{e^2 n_r \omega^3}{3\varepsilon_0 \pi c^3 \hbar} |\langle \psi_i|\hat{r}|\psi_f\rangle|^2 \sum_m p_m \sum_n |\langle \chi_{im}|\chi_{fn}\rangle|^2 \delta(E_{ZPL} + \hbar\omega_{im} - \hbar\omega_{fn} - \hbar\omega) \quad (15),$$

where $e$ is the elementary charge, $n_r$ is the refractive index of the bulk material, $\varepsilon_0$ is the vacuum permittivity, $\hbar\omega$ is the corresponding energy of the emitted photon. $\langle\psi_i|\hat{r}|\psi_f\rangle$ is the transition dipole moment between the band edge state and defect state calculated at Γ point; $p_m$ is the Boltzmann occupation of the initial vibrational state $m$ changing with temperature; $\chi_{im}$ and $\chi_{fn}$ are the vibrational wavefunctions of the initial and final states, and $\hbar\omega_{im}$ and $\hbar\omega_{fn}$ are the corresponding eigenvalues. $E_{ZPL}$ is corresponding to the charge-state transition level calculated in DEC module. At T=0 K, all the charge carriers fall into the lowest-energy vibrational mode of initial state ($m$=0), since there is no thermal activation for excited-state carriers. As a result, one can simply omit the summation over $m$ in Eq. (15) to obtain the result for T=0 K.

In order to evaluate the overlap integral in Eq. (15), one-dimensional configuration coordinate diagram is used with an example given in Fig. 6(b), which adopts a single effective phonon mode to represent all 3N phonons in the system.[49] This phonon mode can be depicted by the distortion of the defects between two charge states; the structural difference $\Delta Q$ in the configuration coordinate can be given by,[50]

$$\Delta Q = \sqrt{\sum_\alpha m_\alpha (\mathbf{R}_{i\alpha} - \mathbf{R}_{f\alpha})^2} \quad (16),$$

where $\mathbf{R}_{i\alpha}$ and $\mathbf{R}_{f\alpha}$ are the Cartesian coordinates of the atom $\alpha$ in initial and final supercell structures, and $m_\alpha$ is its mass. Using configuration coordinate diagram, the effective phonon frequencies $\omega_{im}$ and $\omega_{fn}$ can be obtained by fitting the potential energy surfaces, and the integral can also be numerically calculated.

Configuration coordinate diagram is also useful for further calculating the radiative and nonradiative carrier capture coefficient of such process, which requires the exact calculation of transition dipole moment and electron-phonon coupling matrix element. In the current version, DASP only supports the one-dimensional static coupling method for calculating nonradiative carrier capture;[30] while the original static coupling theory including all 3N phonon modes will be implemented in the future version.[31, 51]

## 3. Calculation Examples

In the following, we will show three examples about the application of DASP in the calculation the defect and dopant properties of semiconductors, including the intrinsic point defect properties of the benchmark system GaN, the intrinsic point defect properties of the low-symmetry quasi-one-dimensional SbSeI, and the PL spectrum of C-doped GaN.

### 3.1 Intrinsic Defects of GaN

GaN is a well-studied wide-band-gap semiconductor whose intrinsic defects have been studied by many groups since 1990s, so it is a good benchmark system for testing our DASP software. In Fig. 7, we show the calculated formation energies of vacancies, antisites and interstitials in GaN as functions of Fermi level, which are well consistent with the published results calculated using the hybrid functional.[52-54]

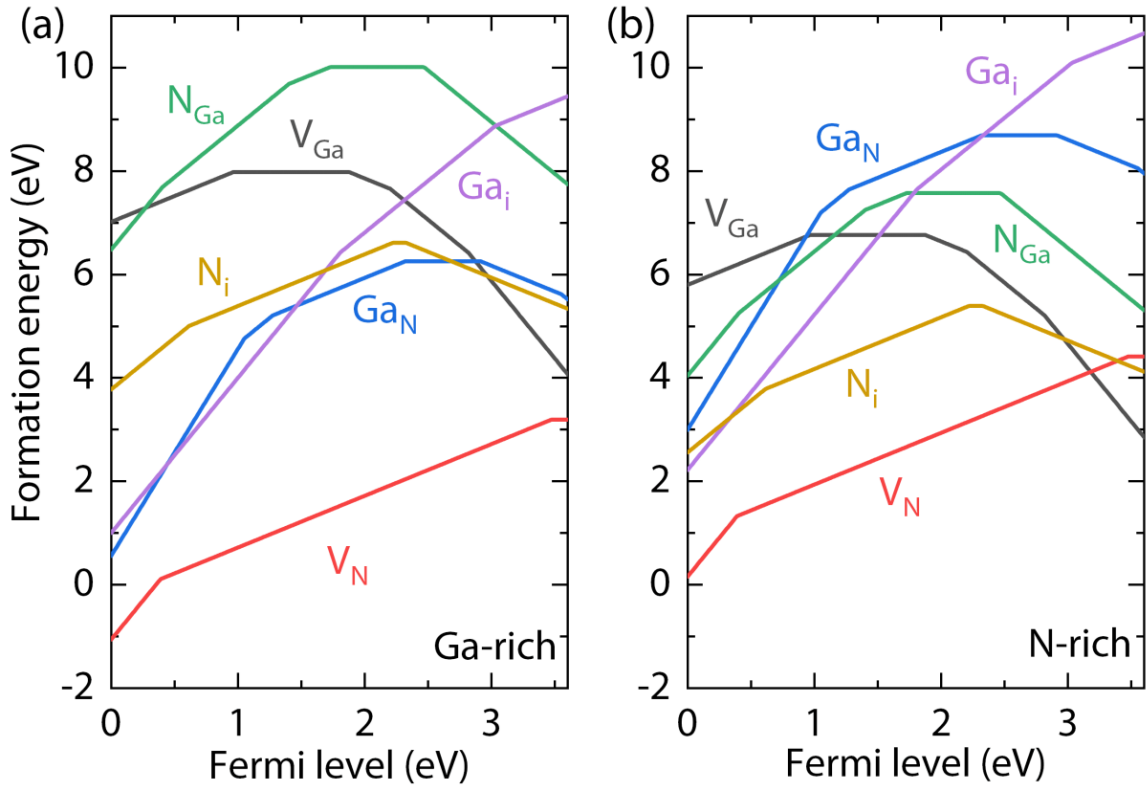

Figure 7. Formation energies of point defects in GaN as functions of Fermi level under (a) Ga-rich and (b) N-rich conditions.[54, 55]

The formation energies of these point defects in neutral states are relatively high in GaN, no matter under Ga-rich or N-rich condition. Among them, the donor defect, N vacancy $V_N$, has the lowest formation energy, but it still cannot produce a high density of electron carriers or good n-type conductivity and shift the Fermi level to close to the CBM level, even under the Ga-rich condition. Therefore, the formation of intrinsic point defects should not produce good n-type conductivity in GaN. For $V_N$, the DASP calculations based on the spin-polarized and hybrid functional ab-initio calculations show a (+/3+) transition level lying 0.4 eV above the VBM, and a (0/+) level close to the CBM. The (+/3+) transition level was usually not found in the early calculations based on the LDA or GGA to the exchange correlation functional,[8, 56] but was found in the recent hybrid functional calculations.[52-54]

## 3.2 Intrinsic Defects of SbSeI

Antimony selenoiodide (SbSeI) has a quasi-one-dimensional structure, as shown in Fig. 8(a), which is similar to that of $Sb_2Se_3$.[24] Its band structure is shown in Fig. 8(b), and has an indirect band gap of 1.79 eV. Its VBM is composed of the hybridized states of Sb-5s, Se-4p and I-5p, while the CBM is mainly Sb-5p. Using the TSC module, two secondary compounds, $SbI_3$ and $Sb_2Se_3$, are found to determine the allowed range of the Sb, Se and I elemental chemical potentials, together with the elemental phases of Sb, Se and I. Fig. 8(c, d) presents the 3D and projected-2D phase stability diagrams in the elemental chemical potential space based on the results of the TSC calculations. The ternary compound SbSeI is stable only in the orange region surrounded by the competing phases $Sb_2Se_3$, $SbI_3$, Sb and Se. The four boundary points, *A*, *B*, *C* and *D*, are selected as the chemical potential conditions for the following DEC and DDC calculations.

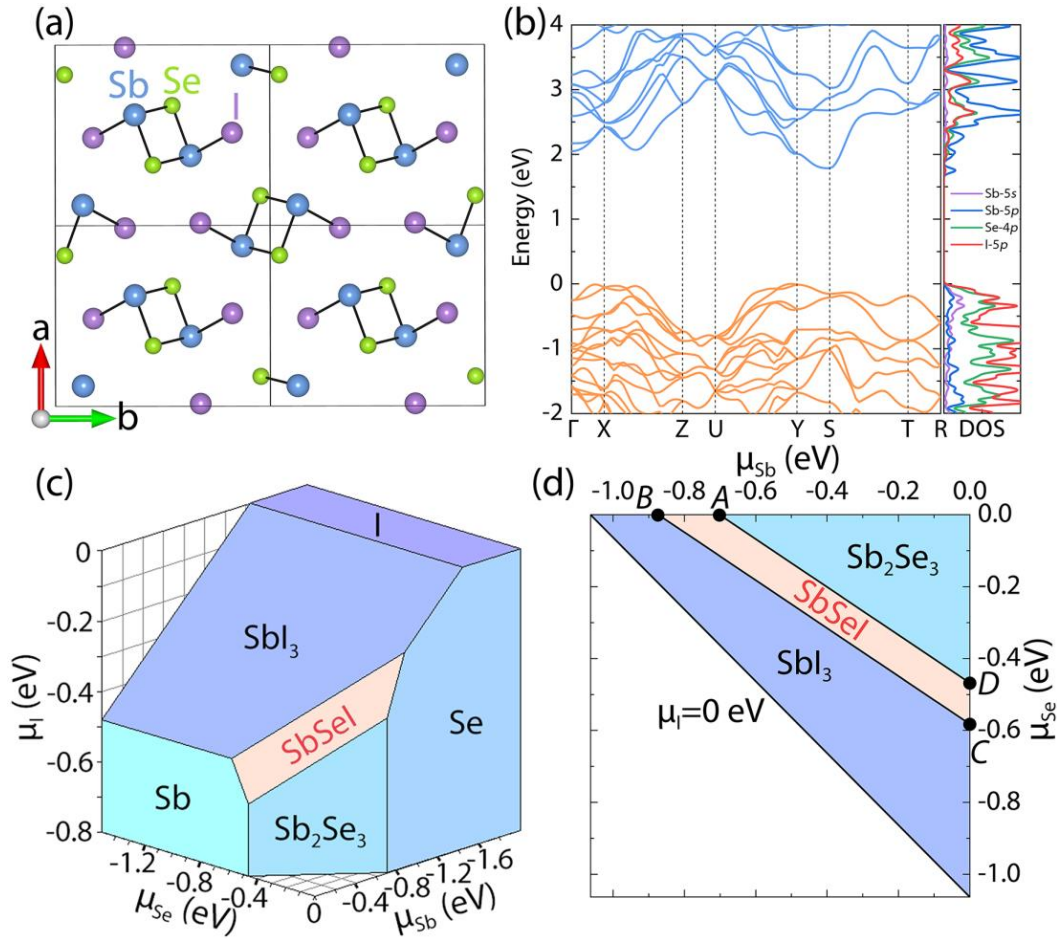

Figure 8. (a) Crystal structure (b) band structure and density of states of SbSeI. (c) 3D and (d) projected-2D plot of phase stability diagram of SbSeI in the chemical potential space.

Fig. 9 shows the calculated formation energies of defects in SbSeI as functions of Fermi level for the four chemical potential conditions, *A*, *B*, *C* and *D*. The densities of these defects and carriers are also calculated using the DDC module and the results for the high-density defects are plotted in Fig. 10(a) as functions of the elemental chemical potential. Meanwhile, the change of Fermi level and carrier densities with the elemental chemical potential conditions is also plotted in Fig. 10(a).

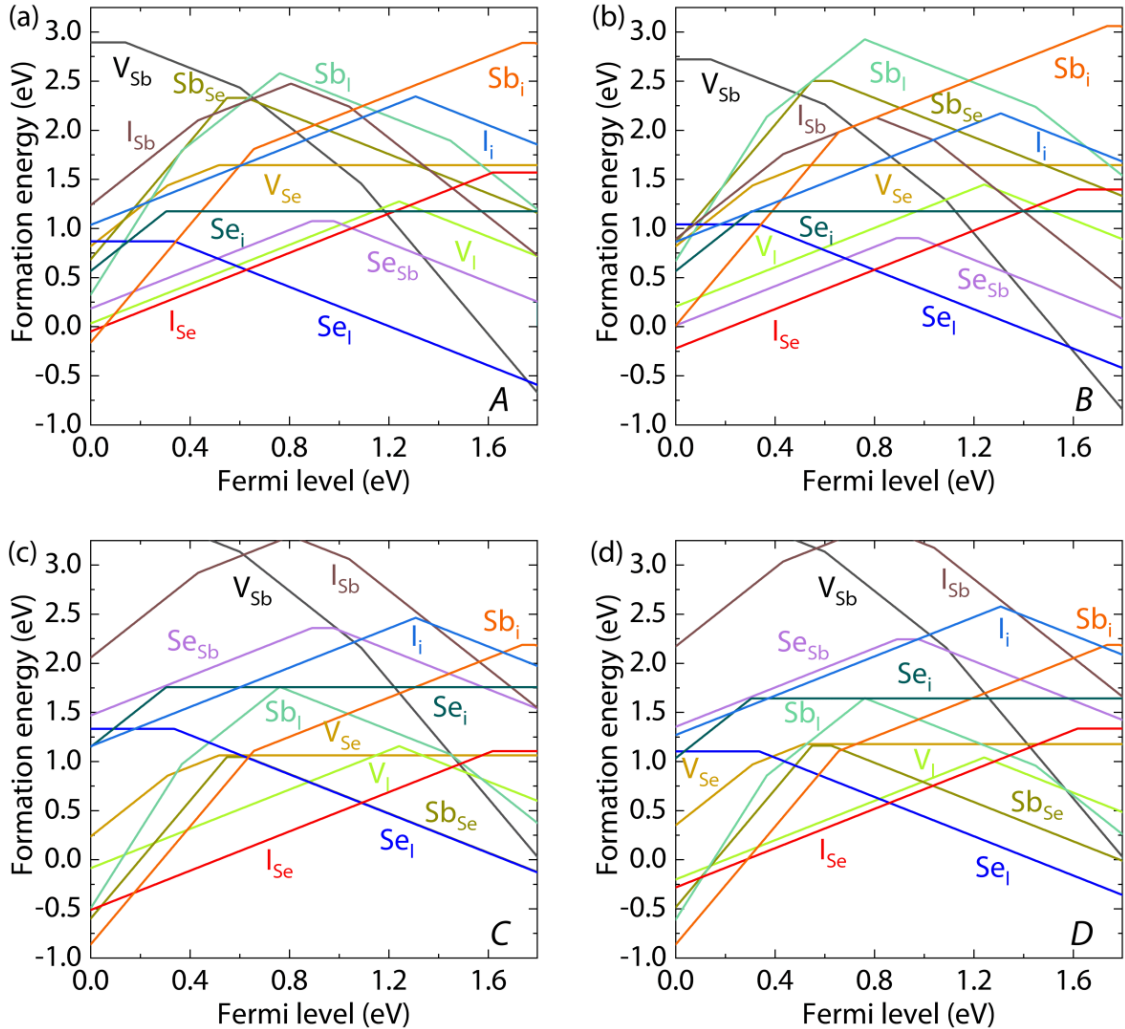

Figure 9. Formation energies of point defects in SbSeI as functions of Fermi level under the chemical potential conditions (a) A, (b) B, (c) C, and (d) D.

It is obvious in Fig. 9 that there are many intrinsic defects with formation energies lower than 2 eV, so the formation of defects should be energetically easy in the quasi-one-dimensional lattice of SbSeI. Among them, the antisite defects between two anions, $Se_I$ and $I_{Se}$, should be the dominant defects with the lowest energy and highest density. $Se_I$ is an acceptor with a (0/–) level at 0.34 eV above VBM, which is not deep. $I_{Se}$ is a donor with a (0/+) level at 0.17 eV below CBM, which is relatively shallow. Fig. 10(b, c) plots the squared wavefunction of the $Se_I$ and $I_{Se}$ defect states, which show that the defect states are actually localized, although their corresponding transition levels are shallow, indicating the supercell size is sufficient for describing these defects in SbSeI. Besides them, there are also several other defects with formation energies lower than 1 eV and deep transition levels in the band gap, *e.g.*, $Se_{Sb}$ and $V_I$, which are both amphoteric with deep (0/+) and (0/–) transition levels.

Although the two dominant defects, the $Se_I$ acceptor and the $I_{Se}$ donor, have quite high density ($10^{17}$-$10^{18}$ cm$^{-3}$) and relatively shallow transition levels (should produce high density of electron and hole carriers), the final equilibrium density of carriers is not high due to the donor-acceptor compensation. Therefore, the Fermi level of the SbSeI sample with high densities of defects is always located near the middle of the band gap, *i.e.*, at least 0.4 eV far from the CBM or VBM level as shown in Fig. 11(a), and the electrical conductivity can only be weakly n-type or weakly p-type, regardless of the elemental chemical potential.

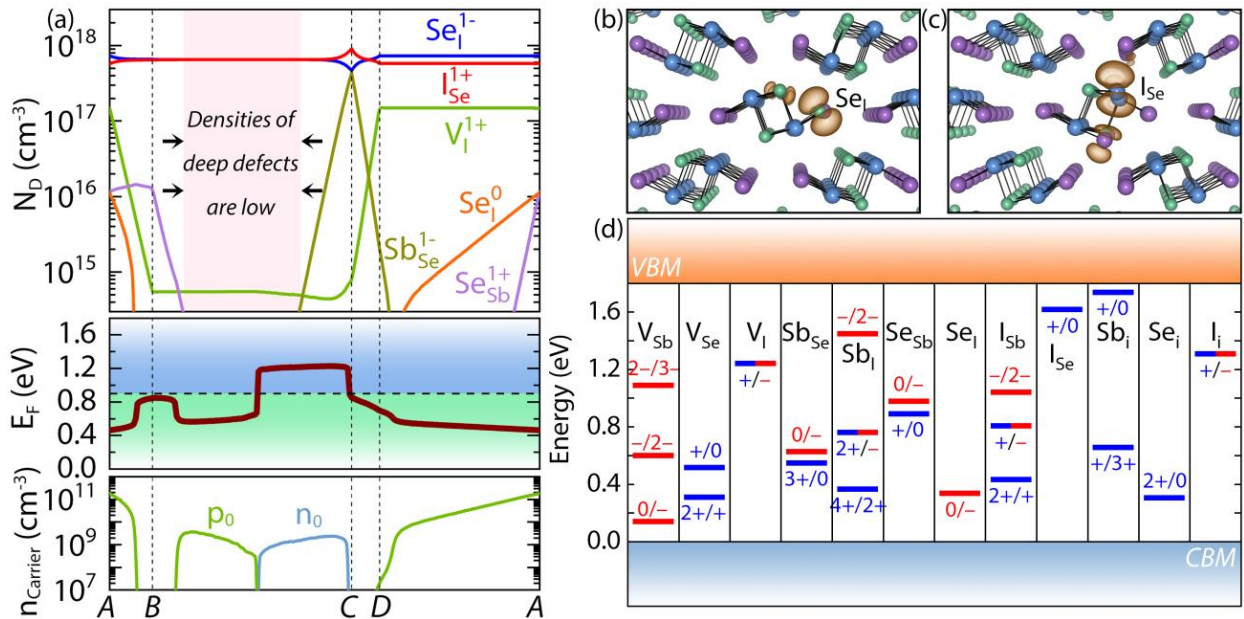

Figure 10. (a) The densities of defects in different charge states, electron and hole carrier densities and Fermi level as functions of the elemental chemical potentials. (b, c) The norm-squared wavefunctions of the neutral $Se_I$ and $I_{Se}$ defect states. (d) The charge-state transition levels of all defects in SbSeI.

In contrast with the shallow $Se_I$ and $I_{Se}$ whose densities are always high, the densities of deep-level defects $Se_{Sb}$ and $V_I$ change significantly as the elemental chemical potentials change. When the chemical potential points are close to the A-D line, the density of $V_I$ is very high; when the points are close to B-C, its density decreases to lower than $10^{15}$ cm$^{-3}$. For $Se_{Sb}$, its density is very high when the chemical potential point is near the C point, but decreases quickly as the point moves away from the C point. These results suggest that the chemical potential should be controlled at the intermediate region of the B-C path, in order to suppress the formation of deep-level defects such as $Se_{Sb}$ and $V_I$.

As we can see, Figs. (8-10) give a good example about the application of DASP for calculating the elemental chemical potential region that stabilizes the compound semiconductors, then the formation energies and transition energy levels of all intrinsic point defects, and finally the densities of defects and electron/hole carriers. For new semiconductors, the similar calculations should be performed to have a complete picture about their defect and dopant properties.

### 3.3 PL Spectrum of C-Doped GaN

The PL spectrum is a widely used optical characterization method of defects in semiconductors. Usually the origin of the PL peaks is attributed to different defects according to the energy differences between the defect level and band edge levels. However, such kind of attribution is questionable, especially when there are several defects whose energy levels are nearly degenerate. DASP can calculate the PL lineshape of different defects, which provide an extra criterion for the identification of the defect origin of PL peaks.

Here we show an example about the calculated PL lineshape of the $C_N$ dopant in GaN. Using DEC module, we find $C_N$ has a (0/+) transition level of 0.39 eV, and a (0/–) transition level of 1.07 eV above the VBM level, which agree with the calculations of Lyons *et al.*[57]. Therefore, the photo-excited electron carrier can be captured by the (0/–) level of $C_N$ through a radiative transition, from the initial state of neutral $C_N$ ($C_N^0$) with an electron on the CBM to the final state of negatively charged $C_N$ ($C_N^-$) with the electron on the (0/–) defect level. This transition is plotted in the configuration coordinate diagram in Fig. 11(b). The direct optical transition level is calculated to be 2.11 eV. The PL lineshape of the transition calculated using the CDC module is plotted in Fig.

11(c), which is in good agreement with experimental measurement and previous calculated result.[50]

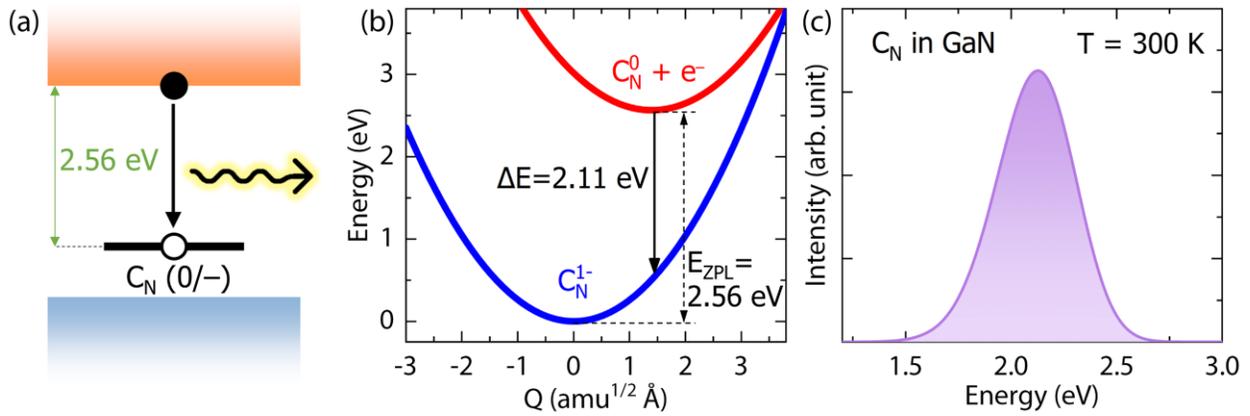

Figure 11. (a) The location of (0/–) transition level of $C_N$ in the band gap of GaN. (b) Configuration coordinate diagram of the radiative transition of an electron from the CBM level to the (0/–) level of $C_N$. (c) Calculated PL lineshape of $C_N$ at T=300 K.

Further CDC calculation shows the radiative carrier capture coefficient $C_n$ is $1.7 \times 10^{-13}$ cm$^3 \cdot$s$^{-1}$, slightly larger than the previous calculated result of $0.7 \times 10^{-13}$ cm$^3 \cdot$s$^{-1}$.[58] The main reason is attributed to the larger momentum matrix element that we calculated, which is $|p_{if}|^2$=0.06 a.u. between the initial and final states.

## 4. Conclusions

We developed a software DASP composed of four modules, TSC, DEC, DDC and CDC, for the automated calculation of defect and dopant properties in the crystalline semiconductors (insulators). DASP just needs the input of the crystal structure file of the semiconductor, then it can visit the materials genome database and call the ab-initio DFT software such as VASP to calculate the defect and dopant properties, including, (i) the chemical potential range of component elements that stabilizes the compound semiconductor (a descriptor of the thermodynamic stability of the compound) and the highest allowed chemical potential of dopant elements; (ii) the formation energies of defects and dopants in different charge states, as functions of elemental chemical potentials and Fermi level, and their charge-state transition energy levels; (iii) the equilibrium densities of defects and dopants, Fermi level, densities of electron and hole carriers, as functions of elemental chemical potentials at growth and working temperature; and (iv) the carrier capture cross sections, radiative and non-radiative carrier recombination rate and lifetime, and the PL spectra. DASP is designed to be an automatic toolbox for the theoretical calculation studies on defects and dopants in semiconductors, and can be used not only for interpreting the electrical and optical characterization experiments of defects and dopants, but also for the quantitative defect and dopant engineering in functional semiconductor devices.

ACKNOWLEDGMENT

We thank Profs. Su-Huai Wei, Xin-Gao Gong, Aron Walsh, Lin-Wang Wang and Yu-Ning Wu, and Drs. Zhen-Kun Yuan, Ji-Qiang Li, Zenghua Cai and Tao Zhang for their long-term collaboration and very helpful discussion.

**References**

[1] Sokrates T. Pantelides, The electronic structure of impurities and other point defects in semiconductors, Reviews of Modern Physics 50, 797-858 (1978).

[2] Ji Sang Park, Sunghyun Kim, Zijuan Xie, Aron Walsh, Point defect engineering in thin-film solar cells, Nature Reviews Materials 3, 194-210 (2018).

[3] R. Ramprasad, H. Zhu, Patrick Rinke, Matthias Scheffler, New Perspective on Formation Energies and Energy Levels of Point Defects in Nonmetals, Physical Review Letters 108, 066404 (2012).

[4] Shiyou Chen, Aron Walsh, Xin-Gao Gong, Su-Huai Wei, Classification of Lattice Defects in the Kesterite $Cu_2ZnSnS_4$ and $Cu_2ZnSnSe_4$ Earth-Abundant Solar Cell Absorbers, Advanced Materials 25, 1522-1539 (2013).

[5] Audrius Alkauskas, Matthew D. McCluskey, Chris G. Van de Walle, Tutorial: Defects in semiconductors—Combining experiment and theory, Journal of Applied Physics 119, 181101 (2016).

[6] S. B. Zhang, John E. Northrup, Chemical potential dependence of defect formation energies in GaAs: Application to Ga self-diffusion, Physical Review Letters 67, 2339-2342 (1991).

[7] S. B. Zhang, Su-Huai Wei, Alex Zunger, H. Katayama-Yoshida, Defect physics of the $CuInSe_2$ chalcopyrite semiconductor, Physical Review B 57, 9642-9656 (1998).

[8] Jörg Neugebauer, Chris G. Van de Walle, Atomic geometry and electronic structure of native defects in GaN, Physical Review B 50, 8067-8070 (1994).

[9] Chris G. Van de Walle, Jörg Neugebauer, First-principles calculations for defects and impurities: Applications to III-nitrides, Journal of Applied Physics 95, 3851-3879 (2004).

[10] Ji-Hui Yang, Wan-Jian Yin, Ji-Sang Park, Jie Ma, Su-Huai Wei, Review on first-principles study of defect properties of CdTe as a solar cell absorber, Semiconductor Science and Technology 31, 083002 (2016).

[11] Wan-Jian Yin, Tingting Shi, Yanfa Yan, Unusual defect physics in CH3NH3PbI3 perovskite solar cell absorber, Applied Physics Letters 104, 063903 (2014).

[12] Stephan Lany, Alex Zunger, Assessment of correction methods for the band-gap problem and for finite-size effects in supercell defect calculations: Case studies for ZnO and GaAs, Physical Review B 78, 235104 (2008).

[13] D. M. Ceperley, B. J. Alder, Ground state of the electron gas by a stochastic method, Physical Review Letters 45, 566-569 (1980).

[14] John P. Perdew, Kieron Burke, Matthias Ernzerhof, Generalized Gradient Approximation Made Simple, Physical Review Letters 77, 3865-3868 (1996).

[15] David Segev, Anderson Janotti, Chris G. Van de Walle, Self-consistent band-gap corrections in density functional theory using modified pseudopotentials, Physical Review B 75, 035201 (2007).

[16] Audrius Alkauskas, Peter Broqvist, Alfredo Pasquarello, Defect Energy Levels in Density Functional Calculations: Alignment and Band Gap Problem, Physical Review Letters 101, 046405 (2008).

[17] Adisak Boonchun, Walter R. L. Lambrecht, Critical evaluation of the LDA + U approach for band gap corrections in point defect calculations: The oxygen vacancy in ZnO case study, physica status solidi (b) 248, 1043-1051 (2011).

[18] R. Saniz, Y. Xu, M. Matsubara, M. N. Amini, H. Dixit, D. Lamoen, B. Partoens, A simplified approach to the band gap correction of defect formation energies: Al, Ga, and In-doped ZnO, Journal of Physics and Chemistry of Solids 74, 45-50 (2013).

[19] Peter Deák, Bálint Aradi, Thomas Frauenheim, Erik Janzén, Adam Gali, Accurate defect levels obtained from the HSE06 range-separated hybrid functional, Physical Review B 81, 153203 (2010).

[20] Audrius Alkauskas, Peter Broqvist, Alfredo Pasquarello, Defect levels through hybrid density functionals: Insights and applications, physica status solidi (b) 248, 775-789 (2011).

[21] Hannu-Pekka Komsa, Alfredo Pasquarello, Assessing the accuracy of hybrid functionals in the determination of defect levels: Application to the As antisite in GaAs, Physical Review B 84, 075207 (2011).

[22] Stephan Lany, Alex Zunger, Accurate prediction of defect properties in density functional supercell calculations, Modelling and Simulation in Materials Science and Engineering 17, 084002 (2009).

[23] Christoph Freysoldt, Jörg Neugebauer, Chris G. Van de Walle, Fully Ab Initio Finite-Size Corrections for Charged-Defect Supercell Calculations, Physical Review Letters 102, 016402 (2009).

[24] Menglin Huang, Zenghua Cai, Shanshan Wang, Xin-Gao Gong, Su-Huai Wei, Shiyou Chen, More Se Vacancies in Sb2Se3 under Se-Rich Conditions: An Abnormal Behavior Induced by Defect-Correlation in Compensated Compound Semiconductors, Small 17, 2102429 (2021).

[25] Christoph Freysoldt, Blazej Grabowski, Tilmann Hickel, Jörg Neugebauer, Georg Kresse, Anderson Janotti, Chris G. Van de Walle, First-principles calculations for point defects in solids, Reviews of Modern Physics 86, 253-305 (2014).

[26] Su-Huai Wei, Overcoming the doping bottleneck in semiconductors, Comput. Mater. Sci. 30, 337-348 (2004).

[27] Anubhav Jain, Shyue Ping Ong, Geoffroy Hautier, Wei Chen, William Davidson Richards, Stephen Dacek, Shreyas Cholia, Dan Gunter, David Skinner, Gerbrand Ceder, Kristin A. Persson, Commentary: The Materials Project: A materials genome approach to accelerating materials innovation, APL Materials 1, 011002 (2013).

[28] G. Kresse, J. Furthmüller, Efficient iterative schemes for ab initio total-energy calculations using a plane-wave basis set, Physical Review B 54, 11169-11186 (1996).

[29] Paolo Giannozzi, Stefano Baroni, Nicola Bonini, Matteo Calandra, Roberto Car, Carlo Cavazzoni, Davide Ceresoli, Guido L. Chiarotti, Matteo Cococcioni, Ismaila Dabo, Andrea Dal Corso, Stefano de Gironcoli, Stefano Fabris, Guido Fratesi, Ralph Gebauer, Uwe Gerstmann, Christos Gougoussis, Anton Kokalj, Michele Lazzeri, Layla Martin-Samos, Nicola Marzari, Francesco Mauri, Riccardo Mazzarello, Stefano Paolini, Alfredo Pasquarello, Lorenzo Paulatto, Carlo Sbraccia, Sandro Scandolo, Gabriele Sclauzero, Ari P. Seitsonen, Alexander Smogunov, Paolo Umari, Renata M. Wentzcovitch, QUANTUM ESPRESSO: a modular and open-source software project for quantum simulations of materials, Journal of Physics: Condensed Matter 21, 395502 (2009).

[30] Audrius Alkauskas, Qimin Yan, Chris G. Van de Walle, First-principles theory of nonradiative carrier capture via multiphonon emission, Physical Review B 90, 075202 (2014).

[31] Lin Shi, Ke Xu, Lin-Wang Wang, Comparative study of ab initio nonradiative recombination rate calculations under different formalisms, Physical Review B 91, 205315 (2015).

[32] Chris G. Van de Walle, DX-center formation in wurtzite and zinc-blende $Al_xGa_{1-x}N$, Physical Review B 57, R2033-R2036 (1998).

[33] C. H. Park, S. B. Zhang, Su-Huai Wei, Origin of p-type doping difficulty in ZnO: The impurity perspective, Physical Review B 66, 073202 (2002).

[34] Jingbo Li, Su-Huai Wei, Lin-Wang Wang, Stability of the $DX^-$ Center in GaAs Quantum Dots, Physical Review Letters 94, 185501 (2005).

[35] Jin-Ling Li, Jingxiu Yang, Tom Wu, Su-Huai Wei, Formation of DY center as n-type limiting defects in octahedral semiconductors: the case of Bi-doped hybrid halide perovskites, Journal of Materials Chemistry C 7, 4230-4234 (2019).

[36] Hannu-Pekka Komsa, Arkady V. Krasheninnikov, Native defects in bulk and monolayer $MoS_2$ from first principles, Physical Review B 91, 125304 (2015).

[37] Yu Kumagai, Fumiyasu Oba, Electrostatics-based finite-size corrections for first-principles point defect calculations, Physical Review B 89, 195205 (2014).

[38] Dan Wang, Dong Han, Xian-Bin Li, Sheng-Yi Xie, Nian-Ke Chen, Wei Quan Tian, Damien West, Hong-Bo Sun, S. B Zhang, Determination of Formation and Ionization Energies of Charged Defects in Two-Dimensional Materials, Physical Review Letters 114, 196801 (2015).

[39] Hannu-Pekka Komsa, Alfredo Pasquarello, Finite-Size Supercell Correction for Charged Defects at Surfaces and Interfaces, Physical Review Letters 110, 095505 (2013).

[40] Yu-Ning Wu, X. G. Zhang, Sokrates T. Pantelides, Fundamental Resolution of Difficulties in the Theory of Charged Point Defects in Semiconductors, Physical Review Letters 119, 105501 (2017).

[41] Jin Xiao, Kaike Yang, Dan Guo, Tao Shen, Hui-Xiong Deng, Shu-Shen Li, Jun-Wei Luo, Su-Huai Wei, Realistic dimension-independent approach for charged-defect calculations in semiconductors, Physical Review B 101, 165306 (2020).

[42] Guo-Jun Zhu, Ji-Hui Yang, Xin-Gao Gong, Self-consistently determining structures of charged defects and defect ionization energies in low-dimensional semiconductors, Physical Review B 102, 035202 (2020).

[43] Christoph Freysoldt, Jörg Neugebauer, First-principles calculations for charged defects at surfaces, interfaces, and two-dimensional materials in the presence of electric fields, Physical Review B 97, 205425 (2018).

[44] Jie Ma, Su-Huai Wei, T. A. Gessert, Ken K. Chin, Carrier density and compensation in semiconductors with multiple dopants and multiple transition energy levels: Case of Cu impurities in CdTe, Physical Review B 83, 245207 (2011).

[45] Ji-Hui Yang, Ji-Sang Park, Joongoo Kang, Wyatt Metzger, Teresa Barnes, Su-Huai Wei, Tuning the Fermi level beyond the equilibrium doping limit through quenching: The case of CdTe, Physical Review B 90, 245202 (2014).

[46] Jinchen Wei, Lilai Jiang, Menglin Huang, Yuning Wu, Shiyou Chen, Intrinsic Defect Limit to the Growth of Orthorhombic $HfO_2$ and $(Hf,Zr)O_2$ with Strong Ferroelectricity: First-Principles Insights, Advanced Functional Materials 31, 2104913 (2021).

[47] Arthur Marshall Stoneham, Theory of defects in solids: electronic structure of defects in insulators and semiconductors, Oxford University Press (2001).

[48] Menglin Huang, Shan-Shan Wang, Yu-Ning Wu, Shiyou Chen, Defect Physics of Ternary Semiconductor $ZnGeP_2$ with a High Density of Anion-Cation Antisites: A First-Principles Study, Physical Review Applied 15, (2021).

[49] F. Schanovsky, W. Gös, T. Grasser, Multiphonon hole trapping from first principles, Journal of Vacuum Science & Technology B 29, 01A201 (2011).

[50] Audrius Alkauskas, John L. Lyons, Daniel Steiauf, Chris G. Van de Walle, First-Principles Calculations of Luminescence Spectrum Line Shapes for Defects in Semiconductors: The Example of GaN and ZnO, Physical Review Letters 109, 267401 (2012).

[51] Lele Cai, Shanshan Wang, Menglin Huang, Yu-Ning Wu, Shiyou Chen, First-principles identification of deep energy levels of sulfur impurities in silicon and their carrier capture cross sections, Journal of Physics D-Applied Physics 54, 335103 (2021).

[52] I. C Diallo, D. O Demchenko, Native Point Defects in GaN: A Hybrid-Functional Study, Physical Review Applied 6, 064002 (2016).

[53] Giacomo Miceli, Alfredo Pasquarello, Energetics of native point defects in GaN: A density-functional study, Microelectronic Engineering 147, 51-54 (2015).

[54] John L. Lyons, Chris G. Van de Walle, Computationally predicted energies and properties of defects in GaN, npj Computational Materials 3, 12 (2017).

[55] He Li, Menglin Huang, Shiyou Chen, First-principles exploration of defect-pairs in GaN, Journal of Semiconductors 41, 032104 (2020).

[56] T. Mattila, R. M. Nieminen, Point-defect complexes and broadband luminescence in GaN and AlN, Physical Review B 55, 9571-9576 (1997).

[57] J. L. Lyons, A. Janotti, C. G. Van de Walle, Effects of carbon on the electrical and optical properties of InN, GaN, and AlN, Physical Review B 89, 035204 (2014).

[58] Cyrus E. Dreyer, Audrius Alkauskas, John L. Lyons, Chris G. Van de Walle, Radiative capture rates at deep defects from electronic structure calculations, Physical Review B 102, 085305 (2020).